\documentclass[amsmath,amssymb, superscriptaddress, preprint]{revtex4-1}

\usepackage[T1]{fontenc}
\usepackage[ansinew]{inputenc}
\usepackage[english]{babel}
\usepackage{amsfonts}
\usepackage{array}
\usepackage{amsthm}

\usepackage{graphicx}

\usepackage{braket}
\usepackage{eucal}
\usepackage{verbatim}
\usepackage[table]{xcolor}
\usepackage{caption}
\usepackage{bm}

\usepackage[pdftex,plainpages=false,colorlinks,hyperindex,bookmarksopen,linkcolor=black,citecolor=blue,urlcolor=blue]{hyperref}

\usepackage{braket}

\newcommand{\be}{\begin{eqnarray}}
\newcommand{\ee}{\end{eqnarray}}

\newcommand{\M}{\mathcal{M}}
\newcommand{\lag}{\mathcal{L}}
\newcommand{\R}{\mathbb{R}}  %%%%% \R = \mathbb{R}.
\newcommand{\N}{\mathbb{N}} %%%% \K = \mathbb{K}.
\def\eg{{\it e.g. }} 
\def\ie{{\it i.e. }}

\begin{document}

\title{Fractional Bosonic Strings}

	\author{Victor Alfonzo Diaz}
		\email{diaz@bo.infn.it}
		\affiliation{Dipartimento di Fisica e Astronomia, Universit\`a di Bologna, Via Irnerio~46, 40126 Bologna, Italy.}	
 		\affiliation{I.N.F.N., Sezione di Bologna, IS - ST$\&$FI, via B.~Pichat~6/2, 40127 Bologna, Italy}
	\author{Andrea Giusti}
		\email{agiusti@bo.infn.it}
		\affiliation{Dipartimento di Fisica e Astronomia, Universit\`a di Bologna, Via Irnerio~46, 40126 Bologna, Italy.}	
 		\affiliation{I.N.F.N., Sezione di Bologna, IS - FLAG, via B.~Pichat~6/2, 40127 Bologna, Italy.}
 		\affiliation{Arnold Sommerfeld Center, Ludwig-Maximilians-Universit\"at, Theresienstra\ss e~37,~80333 M\"unchen, Germany.}

	\date  {\today}%%{January 2016}

\begin{abstract}
The aim of this paper is to present a simple generalization of bosonic string theory in the framework of the theory of fractional variational problems. Specifically, we present a fractional extension of the Polyakov action, for which we compute the general form of the equations of motion and discuss the connection between the new fractional action and a generalization the Nambu-Goto action. Consequently, we analyse the symmetries of the modified Polyakov action and try to fix the gauge, following the classical procedures. Then we solve the equations of motion in a simplified setting. Finally, we present an Hamiltonian description of the classical fractional bosonic string and introduce the fractional light-cone gauge. It is important to remark that, throughout the whole paper, we thoroughly discuss how to recover the known results as an ``integer'' limit of the presented model. 

\vskip 1.0 cm

\begin{center}
\textit{Journal of Mathematical Physics} \textbf{59}, 033509 (2018)\\
DOI: \href{https://doi.org/10.1063/1.5021776}{10.1063/1.5021776}
\end{center}
\end{abstract}

%\keywords{ }

\maketitle

\section{Introduction} \label{sec:intro}
	String theory is one of the very few frameworks that allows for a consistent unification of the gravitational interaction with gauge theories. This is one of the main reasons for which such a theory attracts a great deal of attention in the scientific community resulting to be, in the eyes of many physicists, the leading candidate for describing gravity at the quantum level. On this premise, it is natural to ask a simple question: Can we place all of this on a fractional framework? 
\par 
Fractional calculus has been attracting an increasing interest in both the physical and mathematical community, reaching its peak mostly in the last decade. Indeed, this mathematical approach has been applied in various areas of physics and engineering, see \eg \cite{Baleanu, MainardiB, IC-AG-FM-ZAMP, IC-AG-FM-Bessel, Fabrizio, AG-FCAA-2017, extra, AG-FM_MECC16, JMP-mio, Garra, Silvia-1, Vacaru}. In particular, a peculiar application of such formalism to variational problems \cite{Baleanu, Ata, Almeida, Torres-book} seems to have entered the spotlight. The aim of this section is to briefly review the notion of fractional integral, paying particular attention to its connection with the so called class of fractional variational problems.
	
	From an historical viewpoint, the seed ideas of fractional calculus hides behind the Cauchy formula concerning repeted integrations \cite{Gorenflo-Mainardi, Kilbas}, that reads
	\be \label{eq-cauchy}
	_a J ^n f (t) \equiv \int _a ^t d\tau _n \int _a ^{\tau _n} d\tau _{n-1} \cdots \int _a ^{\tau _2} d\tau _1 \, f(\tau _1) = 
	\frac{1}{(n-1)!} \int _a ^t f(\tau) \, (t - \tau) ^{n-1} \, d\tau \, ,
	\ee
	where $a \in \R$, $t > a$, $n \in \N$ and $f(t)$ is assumed to be, at least, locally absolutely integrable on $t>a$.
	
	From a purely mathematical point of view, the expression of $_a J ^n f (t)$, defined in \eqref{eq-cauchy}, can be readily extended to $_a J ^\alpha f (t)$ for $\alpha \in \R ^+$. Indeed, recalling that $(n-1)! = \Gamma (n)$, $\Gamma (n)$ is the Euler gamma function, and introducing an arbitrary positive real number $\alpha$, one can define the so called \textbf{Riemann-Liouville (fractional) integral of order} $\alpha > 0$,
	\be \label{eq-cauchy}
	_a J ^\alpha f (t) \equiv 
	\frac{1}{\Gamma (\alpha)} \int _a ^t f(\tau) \, (t - \tau) ^{\alpha-1} \, d\tau \, ,
	\ee
	for further details we invite the interested reader to refer to \cite{Gorenflo-Mainardi, Kilbas, MainardiB}. Starting from the classical fundamental theorem of calculus one can easily define the notion of fractional derivative as the \textit{left-inverse} of the fractional integral. Clearly this condition is not enough to guarantee the uniqueness of the definition of such an non-local operator and therefore there exist a plethora of definitions of fractional derivatives, \eg both the Riemann-Liouville and the Caputo fractional derivatives are the left-inverse of \eqref{eq-cauchy}. However, for what it might concern the present work we do not need to introduce any specific generalized notion of derivative and, therefore, we invite the the interested reader to refer to the classical textbooks on the subject, such as \cite{Kilbas}.
\par
	The aim of this paper is to introduce the concept of a fractional time-like parameter in the framework of classical bosonic string theory. The paper is therefore organized as follows:
	
	First, in Section \ref{sec:FC}, we briefly discuss some generalities of fractional variational problems. 
	
	In Section \ref{sec-statement}, we start by fractionalizing the Polyakov action with respect to the time-like parameter on the worldsheet.
 
 	In Section \ref{sec-EOM} we compute the equations of motion and we discuss the connection between the new Polyakov action and an extended version of the Nambu-Goto action.
 
	In Section \ref{sec-symm} and \ref{sec-sol} we discuss the symmetries of the modified Polyakov action and we try to fix the gauge, following the classical procedures used for the ordinary conformal gauge. After that we compute the equations of motion and we we try to solve them in this new gauge. Always, checking the limit in which the fractional parameter $\alpha\rightarrow1$ , where we recover the usual classical bosonic string theory result.
		
	Then, we briefly discuss, in Section \ref{sec-H}, the Hamiltonian description of the classical fractional bosonic string and, after that, in Section \ref{sec-LC} we provide some remarks on a fractional version of the light-cone gauge and we derive the classical mass for a fractional bosonic string.
	 
	Finally, in Section \ref{sec-conclusions}, we present some concluding remarks and some hints for future research.

%%%%%%%%%%%%%%%%
%%%%% FALVA
\section{Fractional Calculus and Variational Problems} \label{sec:FC}
	As widely discussed by Calcagni in \cite{Calcagni-rev, Calcagni-PRL} in order to encode the idea of dimensional flow in the mathematical description of physical systems, one could consider to substitute the Lebesgue measure with the so called Lebesgue-Stieltjes measure, with a peculiar scaling, in the action integral, \ie
	\be 
	S = \int d^D x \, \mathcal{L} \,\,\, \longrightarrow \,\,\, S ^{(\alpha)} = \int d\mu _\alpha (x) \, \mathcal{L} \, ,
	\ee
	where $D \in \N$ is the topological dimension of the $D$-dimensional parameter space of the system, $d \mu _\alpha (x)$ is such that $[\mu _\alpha] = - D \, \alpha$, in momentum units, and $0 < \alpha \leq 1$. This choice is inspired by the fact that all the major quantum gravity theories predict an effective two-dimensional behaviour of physics at very short scales. Hence, physical systems that exhibit an effective change of their dimensionality, as a function of the scale, are likely to show some fractional effects, at least at some level in the dynamical equations. 
	
	For sake of generality we will assume $\alpha = (\alpha _1, \ldots , \alpha _D)$, such that $0 \leq \alpha _i \leq 1$, with $i = 1, \ldots, D$.
	
	A peculiar choice for the modified measure $d \mu _\alpha (x)$, considering an absolutely continuous framework, is given by
	\be \label{eq-modified-measure}
	d \mu _\alpha (x) = d^D x \, v_\alpha (x) \, ,
	\ee
	for some scalar function $v_\alpha (x)$ (see \cite{Calcagni-rev}).

	Now, let $\M$ be the $d$-dimensional \textit{configuration space} and $\Omega \subset \R ^D$, for $D \in \N$, the so called \textit{parameter space}. Denoting with $T \M$ the tangent bundle of $\M$, we can then define a function $\lag$, the \textit{Lagrangian}, such that $\lag \in C^2 (T \M; \R)$. Moreover, denoting with 
	$$\psi \, : \, U \subset \mathcal{M} \, \to \, \psi(U) \subset \R^d$$
	a chart such that $\Phi \in C^2 (\Omega; \M)$ is mapped into $\bm{\Phi} (x) \in \psi(U)$, we can explicitly write the Lagrangian as a function of $\bm{\Phi} (x)$. Now, assuming that $\Phi (x)$ satisfies some fixed Dirichlet boundary conditions on $\partial \Omega$, then the modified action, according to the measure \eqref{eq-modified-measure}, reads
	\be \label{eq-modified-action} 
	S^{(\alpha)} [\bm{\Phi}] = \int _\Omega d^D x \, v_\alpha (x) \, \lag (x, \bm{\Phi}, \partial \bm{\Phi}) \, .
	\ee
	
	The corresponding equations of motion can be easily computed and they read
	\be \label{eq-modifiedEOM}
	\frac{\partial \lag}{\partial \Phi _i (x)} - \partial _k \left( \frac{\partial \lag}{\partial ( \partial _k \Phi _i (x))} \right)
	=
	\frac{\partial _k v_\alpha (x)}{v_\alpha (x)} \, \frac{\partial \lag}{\partial ( \partial _k \Phi _i (x))} \, ,
	\ee	
	where $\partial _k = \partial / \partial x _{k}$, with $k = 1, \ldots, D$, $\Phi _i$ is the $i$-th component of $\bm{\Phi} (x)$, with $i = 1, \ldots, d$. Furthermore, in \eqref{eq-modifiedEOM} we have taken profit of the Einstein's summation convention for repeated indices.
	
	It is now important to remark that the specific choice of the integration measure in \eqref{eq-modified-measure} allows us to connect the modified action principle with the general formalism of fractional calculus. Indeed, as proposed in \cite{El}, we could introduce a specific choice of the function $v_\alpha (x)$, and of the parameter space $\Omega$. In particular, if we choose
	 \be \label{frac-measure}
	_a v_\alpha (\xi ; x) = \prod _{k = 1} ^D \frac{(\xi _k - x_k)^{\alpha _k - 1}}{\Gamma (\alpha _k)} \, , 
	 \quad \Omega _\xi =( a_1, \xi _1 ) \, \times \cdots \times \, ( a_D, \xi _D ) \, ,
	 \ee
	where $\xi = (\xi _1, \ldots, \xi _D)$ and $a = (a _1, \ldots, a _D)$ are two constant real vectors, we have that the corresponding \textbf{Fractional Action} $S^{(\alpha)} [\bm{\Phi}]$ is defined in terms of an integral of the convolution type that represents the $D$-dimensional generalization of the Riemann-Liouville integral, \ie
	\be \label{eq-fractional-action}
	S^{(\alpha)} [\bm{\Phi}] = \int _{\Omega _\xi} d^D x\prod _{k = 1} ^D \frac{(\xi _k - x_k)^{\alpha _k- 1}}{\Gamma (\alpha _k)} \, \lag (x, \bm{\Phi}, \partial \bm{\Phi}) \, .
	\ee
	Thus, given the latter action functional we have that \eqref{eq-modifiedEOM} turns into
	\be \label{eq-Fractional-EOM}
	\frac{\partial \lag}{\partial \Phi _i (x)} - \partial _k \left( \frac{\partial \lag}{\partial ( \partial _k \Phi _i (x))} \right)
	=
	\frac{1 - \alpha _k}{\xi _k - x_k} \, \frac{\partial \lag}{\partial ( \partial _k \Phi _i (x))} \, .
	\ee

	This analysis has found several applications in theoretical physics, in particular it is important to mention the so called \textit{fractional action cosmology model} \cite{Cosmo-1, Cosmo-2}, in which a fractional modification of the Einstein-Hilbert gravitational action is proposed for a Friedmann-Robertson-Walker background.
	
	In the next sections we analyze the implications of the discussed formalism in the classical analysis of bosonic strings, and we will also provide some general comments on the problems arising in the quantization procedure for the resulting \textbf{Fractional Bosonic String theory}.

	It is also worth remarking that an Hamiltonian perspective on theories involving fractional actions was hinted at in \cite{WSEAS}, and could potentially turn out to be useful for future developments of the study presented in this paper.

%%%%%%%%%%%%%%%%%%%
%%%%%%%% Victor's Notes

\section{Statement of the problem} \label{sec-statement}

The propagation of strings on the $d$-dimensional Minkowsky spacetime $\mathbb{R}^{1,d-1}$, usually referred to as the target space, generate a 2-dimensional surface $\Sigma$ called worldsheet (WS) defined by the map
\begin{eqnarray}
\bm{X} \, : \, \Omega \, \longrightarrow \, \mathbb{R}^{1,d-1} \, , \quad (\tau , \, \sigma) \mapsto \bm{X} (\tau,\sigma) = \{ X^\mu (\tau, \sigma)\, , \,\, \quad \mu = 0, 1, \ldots, d-1 \} \, ,  
\end{eqnarray}
called string map, where $\Sigma = \{ X^\mu (\Omega) \}$. Here, $\Omega = \R \times [0, \ell]$, with $\ell > 0$ the length of the string. Now, the most convenient and simple way to describe the dynamics of propagating strings is through the use of the Polyakov action \cite{deser, brink, polyakov, lust, Zwiebach}, 
\begin{eqnarray}\label{Polyact}
S_{P}=-\frac{1}{4\pi\alpha'}\int_{\Omega} d\sigma \, d\tau \,\sqrt{-h}\,h^{ab}(\tau,\sigma)\,\partial_a \bm{X}(\tau,\sigma) \cdot \partial_b \bm{X} (\tau,\sigma) \, ,
\end{eqnarray}
where $a,b=\tau,\sigma$, $h_{ab}(\tau,\sigma)$ is called the intrinsic metric of the WS, $h\equiv \det(h_{ab})$, $\bm{a} \cdot \, \bm{b} = \braket{\bm{a}, \, \bm{b}} _\eta$ with $\braket{\, , \,}_\eta$ the Minkowsky's bilinear form on $\R^{1, d}$ and $\alpha'$ is the Regge slope. It is now important to remark that the dynamical variables $\bm{X}(\tau,\sigma)$ are $d$-vectors from the view point of the target space but they are $d$-scalars from the WS theory. Furthermore, in general the WS  has curvature and therefore it endorse a natural connection given by the Levi-Civita connection of $h_{ab}$. It also endorse a signature $(-,+)$, which explain the minus sign in the square root of the determinant measure in \eqref{Polyact} and the overall minus sign.

The scope of our work is to go a little be further and introduce a generalized version of the Polyakov action defined by
\begin{eqnarray}\label{PolyactFrac}
S^{\alpha, \beta} [h, \bm{X}] = -\frac{1}{4\pi\alpha'} \int _\Omega d\sigma \, d\tau \, v_{\alpha , \beta} (\tau, \sigma) 
\,\sqrt{-h}\,h^{ab}(\tau,\sigma)\,\partial_a \bm{X}(\tau,\sigma) \cdot \partial_b \bm{X} (\tau,\sigma) \, ,
\end{eqnarray}
where $0 \leq \alpha , \beta \leq 1$.

	In particular, for sake of simplicity we will focus on a specific realization of such a generalization. Specifically, we will discuss extensively the case in which
	\be \label{RL-time-measure}
	\Omega = \Omega _t = ( - \infty, t ) \, \times \,  [0, \ell] \, , \quad 
	v_{\alpha , \beta} (x)  = \, _{-\infty} v_{\alpha , 1} (t; \tau, \sigma) = 
	\frac{(t - \tau)^{\alpha - 1}}{\Gamma (\alpha)} \, ,
	\ee 
	where $0 < \alpha < 1$ and $t \in \R$. 
	
	Therefore, in our specific example the modified action reduces to a \textit{time-fractional Polyakov action} given by
	\be \label{our-model}
	S_{\alpha}\equiv-\frac{1}{4\pi\alpha'\,\Gamma(\alpha)}\int_{0}^{\ell}d\sigma\,\int_{-\infty}^{t}(t-\tau)^{\alpha-1}\,d\tau\left[\sqrt{-h}\,h^{ab}(\tau,\sigma)\,\partial_a \bm{X}(\tau,\sigma) \cdot \partial_b \bm{X} (\tau,\sigma) \right].
	\ee

	As we can see in the latter equation, basically we have modified the Polyakov action by changing the integration over $\tau$ with a fractional integral of order $0 < \alpha < 1$ over the same parameter. 
	
	 It is easy to infer that in the limit $\alpha\rightarrow1$ and $t\rightarrow+\infty$ we recover the Polyakov action \eqref{Polyact}. It is clear that these limits do not commute, i.e. we cannot take the $\alpha$-limit after the $t$-limit, otherwise we encounter a divergence in the action.

 \section{Equations of motion} \label{sec-EOM}
	Following the discussion presented in Section \ref{sec:FC} we can easily compute the equations of motion for the general modified Polyakov action \eqref{PolyactFrac}, and we get
	\be \label{eq-motion-gen-1}
	\delta _{\bm{X}} S^{\alpha, \beta} = 0 \,\, \rightsquigarrow \,\, 
	\partial_a \left[ v_{\alpha , \beta} (\tau , \sigma ) \, \sqrt{-h} \, h^{ab} \, \partial_b \bm{X} \right] = 0 \, ,
	\ee
	\be\label{eq-motion-gen-2}	
	\delta _{h} S^{\alpha, \beta} = 0 \,\, \rightsquigarrow \,\, 
	\partial_a \bm{X} \cdot\partial_b \bm{X} - \frac{1}{2}\,h_{ab}\,h^{cd}\,\partial_c \bm{X} \cdot \partial_d \bm{X} = 0 \, .
	\ee

	If we define the energy-momentum tensor of the world-sheet theory in the usual way, \ie
	\be \label{stress-energy}
	T_{ab} \equiv - \frac{4 \pi}{\sqrt{-h}} \, \frac{\delta S^{\alpha , \beta} [h, \bm{X}]}{\delta h ^{a b}} = 
	\frac{1}{\alpha'} \, \left( \partial_a \bm{X} \cdot\partial_b \bm{X} - \frac{1}{2}\,h_{ab}\,h^{cd}\,\partial_c \bm{X}\cdot\partial_d \bm{X} \right) \, ,
	\ee
	and defining the improved d'Alembert operator $\Box _{v}$ as
	\be 
	\Box _{v} \phi \equiv \frac{1}{v_{\alpha , \beta} (\tau , \sigma ) \, \sqrt{-h}} \, 
	\partial_a \left[ v_{\alpha , \beta} (\tau , \sigma ) \, \sqrt{-h} \, h^{ab} \, \partial_b \phi  \right] \, ,
	\ee
	then Eq. \eqref{eq-motion-gen-1} and \eqref{eq-motion-gen-2} read,
	\be
	T_{ab} = 0 \, ,
	\ee
	\be
	\Box _{v} \bm{X} = 0 \, .
	\ee

	It is worth remarking that the vanishing of the stress-energy tensor allows us to connect the modified Polyakov action with a generalized definition of the Nambu-Goto action. Indeed, it is easy to see that the corresponding generalized Nambu-Goto action is given by
	\be 
	S _{NG} ^{\alpha \beta} [\gamma] = - \frac{1}{2 \pi \alpha'} \int _{\Omega _t} d\tau \, d\sigma \, v_{\alpha , \beta} (\tau , \sigma) \, \sqrt{- \gamma} \, ,
	\ee
where $\gamma \equiv \texttt{det} (\gamma _{a b})$ and $\gamma _{a b}$ is the induced metric on the WS, given by $\gamma _{ab} =\partial_a \bm{X}\cdot\partial_b \bm{X}$. As suggested in \cite{Calcagni-rev, Calcagni-PRL, Fractals}, the latter equation tells us that a suitable choice of the measure function $v_{\alpha , \beta} (\tau, \sigma)$, such as the fractional kernel \eqref{frac-measure}, would lead to a geometrical structure of the parameter space $\Omega$ that would resemble, in some sense, the one of a \textit{fractal}. 

	Now, naively we have that the integration over $\sigma$ in the ``ordinary'' bosonic string theory carries an heavy physical significance due to its connection with the fundamental length of the model, \ie the string length. Our aim for this paper is to present a slightly modified description of the evolution of the bosonic string without affecting excessively the physical meaning associated to the fundamental parameter of the theory. Therefore, despite the fact that our model could be carried out in a more general fashion, we will focus our attention to the fractional action defined in \eqref{our-model}. Hence, for the case of our concern the equations of motion reduce to
	\be \label{our-EOM-1}
	T_{ab} = \partial_a \bm{X} \cdot\partial_b \bm{X} -\frac{1}{2}\,h_{ab}\,h^{cd}\,\partial_c \bm{X} \cdot\partial_d \bm{X} = 0 \, ,
	\ee
	\be \label{our-EOM-2}
	\partial_a \left[(t-\tau)^{\alpha-1}\sqrt{-h}\,h^{ab}\,\partial_b \bm{X}  \right] = 0 \, .
	\ee				
As we can see, these are highly non-linear partial differential equations but, due to the symmetries of the action \eqref{our-model}, they can be simplified, as discussed in the following section.	

\section{Symmetries} \label{sec-symm}
	First of all, we need to differentiate two different types of symmetries of the action: the target space symmetries and the WS symmetries.
	
	\subsubsection*{Target space symmetries}
	It is clear from \eqref{our-model}  that our action is completely Poincar\`e invariant, that is the action is invariant under the transformation
	\be 
	\begin{split}
	\bm{X}(\tau,\sigma) &\rightarrow 
	\tilde{\bm{X}}(\tau,\sigma) = \Lambda \, \bm{X}(\tau,\sigma)+\bm{a}\, , \\
	h_{ab}(\tau,\sigma) &\rightarrow
	\tilde{h}_{ab}(\tau,\sigma)=h_{ab}(\tau,\sigma) \, ,
	\end{split}
	\ee	
	where $\Lambda\in SO(1,d-1)$ and $\bm{a}$ is a constant $d$-vector.
	
	\subsubsection*{Worldsheet symmetries}
	We have several symmetries acting on the WS:
	\vskip 0.2 cm
	\noindent First, we have the reparametrization invariance in the $\sigma$-direction, indeed performing the transformation
	\be 
	\begin{split}
	(\tau,\sigma) \,\,\, &\rightarrow \,\,\,
	(\tilde{\tau},\tilde{\sigma})=(\tau,\sigma-\zeta(\tau,\sigma)) \, ,\\
	\bm{X}(\tau,\sigma)	\,\,\, &\rightarrow \,\,\,
	\tilde{\bm{X}}(\tilde{\tau},\tilde{\sigma})=\bm{X}(\tau,\sigma)-\zeta \, \partial_{\sigma} \bm{X}(\tau,\sigma)\, \\
	h_{ab}(\tau,\sigma)	 \,\,\, &\rightarrow \,\,\, 
	\tilde{h}_{ab}(\tilde{\tau},\tilde{\sigma})= 
	h_{ab}(\tau,\sigma)+\zeta \partial_{\sigma}h_{ab}+(\partial_a \zeta)h_{\sigma b}+(\partial_b \zeta)h_{\sigma a}\, ,
	\end{split}
	\ee
	the fractional action \eqref{our-model} remains invariant.
	\vskip 0.2 cm
	\noindent Furthermore, we also have the so called Weyl Invariance, indeed the fractional action \eqref{our-model} is invariant under the transformation
	\be 
	\begin{split}
	\bm{X}(\tau,\sigma) \,\,\, &\rightarrow \,\,\, 
	\tilde{\bm{X}}(\tilde{\tau},\tilde{\sigma})=\bm{X}(\tau,\sigma) \, , \\
	h_{ab}(\tau,\sigma) \,\,\, &\rightarrow \,\,\, 
	\tilde{h}_{ab}(\tilde{\tau},\tilde{\sigma}) = e^{2\rho(\tau,\sigma)}\,h_{ab}(\tau,\sigma).
	\end{split}
	\ee

	\vskip 0.2 cm
	
	Taking profit of the WS symmetries we can reduce the degrees of freedom (d.o.f) of the intrinsic metric $h_{ab}$. In 2 dimension the metric $h_{ab}$ has 3 d.o.f, using the reparametrization in the $\sigma$-direction we can remove one of them and, by means of the Weyl invariance, we can also remove the other one. This leaves us with one d.o.f for the metric. Therefore, without loosing of generality we can write
\begin{eqnarray} \label{our-metric}
h_{ab}=\left(
\begin{array}{cc}
-1 & 0\\
 0 &[f^2(\tau,\sigma)]^{\alpha-1}
\end{array}
\right),
\end{eqnarray}
with $f(\tau,\sigma)$ an arbitrary function and in the limit $\alpha\rightarrow1$, we recover the usual flat metric. Furthermore, we also need to respect the signature of the metric, thus $h_{\sigma\sigma}(\tau,\sigma)>0$. 

	Clearly, if we were to fractionalize the action with respect to both $\tau$ and $\sigma$ we would lose the $\sigma$-reparametrization invariance as well and therefore we would be left with two d.o.f. for the intrinsic metric $h_{ab}$. Hence, we would have that $h_{ab} = \texttt{diag} \left[ f(\tau, \sigma), \, g(\tau, \sigma) \right]$, with $f(\tau, \sigma)$ and $g (\tau, \sigma)$ two arbitrary functions of the WS parameters.
	 
Now, using the metric \eqref{our-metric} the equation of motion can be written as
\be \label{our-EOM-sym-1}
(\alpha-1)f^{2\alpha-2}\left[\frac{\dot{f}}{f}-\frac{1}{t-\tau}\right]\,\dot{\bm{X}}+(\alpha-1)\,\frac{f'}{f}\,\bm{X}'-\bm{X}''+f^{2\alpha-2}\,\ddot{\bm{X}} 
=0 \, ,
\ee
\be \label{our-EOM-sym-2}
|| f^{\alpha-1}\,\dot{\bm{X}}\pm \bm{X}' ||^2 = 0 \, ,
\ee
where the prime represents the derivative with respect to $\sigma$, the dot is understood as the derivative with respect to $\tau$ and $|| \bm{a} ||^2 := \bm{a} \cdot \bm{a}$.

\section{Solutions of the equations of motion} \label{sec-sol}
	
	The next step is try to solve these equations. Notice that Eq.~\eqref{our-EOM-sym-1}, after the gauge fixing, is still a set of highly non-linear equations parametrised by a real function $f(\tau,\sigma)$. Clearly, it is not possible to find a solution in a closed form as a function of $f(\tau,\sigma)$, therefore we can try to simplify the discussion by studying the solutions of Eq.~\eqref{our-EOM-sym-1} for specific choices of this function. 
	
	Here, for sake of simplicity, we will focus only on the very simple case with $f(\tau,\sigma) = 1$, that allows us to diagonalise the intrinsic metric $h_{ab}$. Hence, Eq.~\eqref{our-EOM-sym-1} and \eqref{our-EOM-sym-2} now read
	\be \label{EOM-f-1}
	\ddot{\bm{X}} - {\bm{X}''}  = \frac{\alpha - 1}{t - \tau} \, \dot{\bm{X}} \, ,	
	\ee
	\be \label{Constraint-f-1}
	|| \dot{\bm{X}}\pm \bm{X}' ||^2 = 0 \, ,
	\ee
	which clearly shows that the introduction of the time-fractional Riemann-Liouville measure \eqref{RL-time-measure} leads to an extra term that resembles the contribution of fluid viscosity in classical fluid dynamics.
	
	Moreover, to further simplify the discussion, in the following we will consider the case with $\ell = 2 \, \pi$ together with periodic boundary conditions
	$$ \bm{X} (\tau , \sigma + 2 \, \pi) = \bm{X} (\tau , \sigma) \, , \,\,\, \forall \tau \in (-\infty , t) \, ,   $$
	\ie we consider the case of a \textit{closed string}.
	
	It is important to remark that the components of $\bm{X}$ do not mix with one another in the system given in Eq.~\eqref{EOM-f-1}, therefore the problem of finding the solutions of this system of partial differential equations reduces to the problem of finding the solutions, compatible with the constraints and with the cyclic boundary conditions, for the following nonlinear partial differential equation
	\be \label{NODE}
	\ddot{X} - X'' = \frac{\alpha - 1}{t - \tau} \, \dot{X} \, ,
	\ee
	with $X = X (\tau, \sigma)$ a real function on the WS.
	
	If we tackle this problem by means of the method of separation of variables, \ie assuming an ansatz of the form $X (\tau, \sigma) = T (\tau) \, \Sigma ( \sigma )$, we get that Eq.~\eqref{NODE} rewrites as
	\be \label{SepVar}
	\frac{\ddot{T}}{T} - \frac{\alpha - 1}{t - \tau}\frac{\dot{T}}{T} = \frac{\Sigma''}{\Sigma} \, .
	\ee
	Since the right hand side depends only on $\sigma$ and the left hand side only on $\tau$, both sides are equal to some constant value $\lambda \in \R$. Thus,
	\be \label{eqSigma}
	\Sigma'' =  \lambda \, \Sigma \, ,
	\ee
	\be \label{eqT}
	\ddot{T} - \frac{\alpha - 1}{t - \tau} \, \dot{T} - \lambda \, T = 0
	\ee
We shall now compute the solutions of these two ordinary differential equations as functions of $\lambda$. To do that we can actually distinguish three cases, namely: $\lambda < 0$, $\lambda = 0$ and $\lambda > 0$.

\vskip 0.3 cm  	
  	
	\noindent \textbf{Case} $\lambda > 0$:
	
	It is trivial to see that \eqref{eqSigma} is solved by
	$$ 
	\Sigma (\sigma) = a_1 \, e^{\sqrt{\lambda} \, \sigma} + a_2 \, e^{- \sqrt{\lambda} \, \sigma} \, , \quad
	a_1 \, , \, a_2 \in \R	 
	$$
	which is inconsistent with the periodic boundary conditions. Therefore, there are no solutions of \eqref{NODE} for $\lambda > 0$.
	
	\vskip 0.3 cm  	
  	
	\noindent \textbf{Case} $\lambda = 0$:
	
	In this case, Eq.~\eqref{eqSigma} reads
	$$
		\Sigma'' = 0 \, ,
	$$
	whose fundamental solution is given by
	$$ 
	\Sigma (\sigma) = a \,  \sigma + C \, , \quad a, C \in \R \, .
	$$
	If we now impose the periodic boundary conditions, we get
	\be 
	\Sigma _{\lambda = 0} (\sigma) = C \, , \quad C \in \R \, .
	\ee
	Besides, Eq.~\eqref{eqT} turns into
	$$
		\ddot{T} - \frac{\alpha - 1}{t - \tau} \, \dot{T} = 0 \, .
	$$
	Setting $z = t - \tau$ and recalling that $\partial _\tau = - \partial _z$, the latter can be rewritten as 
	$$
	\frac{d^2 T}{dz ^2} + \frac{\alpha - 1}{z} \, \frac{d T}{dz} = 0 \, .
	$$ 
	This equation is now trivially solved by
	\be
	T _{\lambda = 0} (z) = \frac{f _1}{2 - \alpha} \, z^{2 - \alpha} + f_0 \, , \quad f_0, f_1 \in \R \, , 
	\ee
	or equivalently,
	\be 
	T _{\lambda = 0} (\tau) = \frac{f _1}{2 - \alpha} \, (t - \tau)^{2 - \alpha} + f_0 \, , \quad f_0, f_1 \in \R \, .
	\ee
	 
	 \vskip 0.3 cm  	
  	
	\noindent \textbf{Case} $\lambda < 0$:
	
	Setting $\lambda = - m^2$ we have that Eq.~\eqref{eqSigma} now reads
	$$
	\Sigma'' + m^2 \, \Sigma = 0 \, ,
	$$
	which is an harmonic oscillator of frequency $m$ and  whose solutions are given by
	\be
	\Sigma _m (\sigma) \equiv \Sigma _{\lambda = - m^2} (\sigma) = A _m \, e^{i \, m \, \sigma} + B _m \, e^{- i \, m \, \sigma} \, ,
 	\ee
	with $A_m, B_m \in \R$ and $m \in \N$. Indeed, from the periodic boundary condition it is easy to infer that $m$ must be a positive integer.
	
	If we now consider the corresponding equation in the $z = t - \tau$ auxiliary variable, \ie
	$$
	\frac{d^2 T}{dz ^2} + \frac{\alpha - 1}{z} \, \frac{d T}{dz} + m^2 \, T (z) = 0 \, ,
	$$
	whose solutions are given by 
	\be
	T _m (z) \equiv T_{\lambda = - m^2} (z) = z ^{\nu} \left[ C_m \, J_{-\nu} (m \, z) + D_m \, Y_{-\nu} (m \, z) \right] 
	\, , \quad \nu = \frac{2 - \alpha}{2} \, ,
	\ee	
where $C_m, D_m \in \R$, $m \in \N$ and $J$ is a Bessel function of the first kind and $Y$ is a Bessel function of the second kind.
	
 \vskip 0.3 cm
 
 Now, putting everything together we have that (in terms of the auxiliary variable $z$ for sake of clarity)
 \be 
 X(z, \sigma) = T_0 (z) \, \Sigma _{0} (\sigma) + \sum _{m = 1} ^\infty T_m (z) \, \Sigma _{m} (\sigma) \, ,
 \ee	
that after some algebraic manipulation turns into
\begin{equation}
 \begin{split}
 X(z, \sigma) = C \, f_0 + C \, f_1 \, \frac{z^{2 \nu}}{2 \nu} + \sum _{m = 1} ^\infty  z ^{\nu} \Big[ &a_m \, J_{-\nu} (m \, z) \, e^{i \, m \, \sigma} + b_m \, Y_{-\nu} (m \, z) \, e^{i \, m \, \sigma} +\\
 & c_m \, J_{-\nu} (m \, z) \, e^{-i \, m \, \sigma} + d_m \, Y_{-\nu} (m \, z) \, e^{-i \, m \, \sigma} \Big] \, .
\end{split}
\end{equation}
If we then recall the connection between the Bessel functions $J$ and $Y$ with the Hankel functions $H^{(1)}$ and $H^{(2)}$ \cite{Abram-Steg}, \ie
\be 
H^{(1)} _\alpha (x) = J_\alpha (x) + i \, Y _\alpha (x) \, ,
\ee 
\be 
H^{(2)} _\alpha (x) = J_\alpha (x) - i \, Y _\alpha (x) \, ,
\ee
and that 
$$
[H^{(1)} _{-\nu} (x)]^\ast = H^{(2)} _{-\nu} (x) \, , \quad [H^{(2)} _{-\nu} (x)]^\ast = H^{(1)} _{-\nu} (x)
$$ 
provided that $x \in \R$ and $1/2 < \nu < 1$, then imposing the reality condition, and with some simple manipulations of the expansion coefficients, we get
\begin{equation} \label{expanded}
\begin{split}
 X(z, \sigma) = X_0 - \frac{\sqrt{2 \, \alpha'} \, z^{2 \nu}}{2 \nu} \, \alpha _0
  - i \, \sqrt{\frac{\alpha' \pi}{4}} \sum _{m = 1} ^\infty  \frac{z ^{\nu}}{\sqrt{m}} \Big[ &( \alpha _m \, e^{i \, m\, \sigma} + \widetilde{\alpha} _m \, e^{- i \, m\, \sigma} ) \, H^{(2)} _{-\nu} (mz) \\
  &- ( \alpha _m ^\ast \, e^{-i \, m\, \sigma} + \widetilde{\alpha} ^\ast _m \, e^{i \, m\, \sigma} ) \, H^{(1)} _{-\nu} (mz)  \Big] \, ,
\end{split}
\end{equation}
where $\ast$ denotes the complex conjugate.

	If we then introduce a function $\mathcal{E} (\nu, m \, ; \, z)$ such that
	\begin{equation} \label{new-function}
\mathcal{E} (\nu, m \, ; \, z) := 
\sqrt{\frac{\pi}{2}} \, z^\nu \times
\left\{
\begin{aligned}
& \sqrt{-m} \, H ^{(1)}_{-\nu} (- m z) \, , \,\, m \in \mathbb{Z} ^- \, , \\
& \sqrt{m} \, H ^{(2)}_{-\nu} (m z) \, , \,\, m \in \mathbb{N} \, , \\
 \end{aligned}
\right.
\end{equation}
then, \eqref{expanded} can be expressed in a more compact form, \ie
\begin{equation} \label{compact}
 X (z, \sigma) = X _0 - \frac{\sqrt{2 \, \alpha'} \, z^{2 \nu}}{2 \nu} \, \alpha _0
  - i \, \sqrt{\frac{\alpha'}{2}} \sum _{m \neq 0}  \left( \frac{\alpha _m}{m} \, e^{i \, m\, \sigma} +  \frac{\widetilde{\alpha} _m}{m} \, e^{- i \, m\, \sigma} \right) \, \mathcal{E} (\nu, m \, ; \, z) \, , 
\end{equation}
where $\alpha _m$ and $\widetilde{\alpha} _m$ are now being redefined in a way that preserves the reality condition, namely $\alpha _m ^\ast = \alpha _{-m}$ and $\widetilde{\alpha} _m ^\ast = \widetilde{\alpha} _{-m}$ for all $m \in \mathbb{Z}$.

	The last expression appears to be particularly useful due to the fact that allows us to immediately read off the ``ordinary limit'' of our solution (\ie $z \mapsto -\tau$ and $\alpha \to 1$), indeed recalling that \cite{Abram-Steg}
	\be 
	H^{(1)} _\alpha (x) = \frac{J _{- \alpha} (x) - e^{- \alpha \pi i}\, J _{\alpha} (x)}{i \, \sin (\alpha \pi)} \, ,
	\ee
	\be 
	H^{(2)} _\alpha (x) = \frac{J _{- \alpha} (x) - e^{\alpha \pi i}\, J _{\alpha} (x)}{- i \, \sin (\alpha \pi)} \, ,
	\ee
	hence we get that
	\begin{equation}
\mathcal{E} (\nu, m \, ; \, z) \, \mapsto \, \mathcal{E} \left(1/2, m \, ; \tau \right) = 
\left\{
\begin{aligned}
& e^{i \, m \, \tau} \, , \,\, m \in \mathbb{Z} ^- \, , \\
& e^{i \, m \, \tau} \, , \,\, m \in \mathbb{N} \, , \\
 \end{aligned}
\right.
\end{equation}
where we have taken advantage of the known relations
$$
H ^{(1)}_{-1/2} (z) =  \sqrt{\frac{2}{\pi \, z}} \, e^{i \, z} \, , \quad 
H ^{(2)}_{-1/2} (z) =  \sqrt{\frac{2}{\pi \, z}} \, e^{- i \, z} \, .
$$
	Therefore, Eq.~\eqref{compact} turns into 
	\begin{equation}
 X (z, \sigma) \, \mapsto \, X^{\alpha = 1} (\tau , \sigma) = X_0 - \sqrt{2 \, \alpha'} \, \tau \, \alpha _0
  - i \, \sqrt{\frac{\alpha'}{2}} \sum _{m \neq 0} \left(\frac{\alpha _m}{m} \, e^{i \, m \, \sigma} +  \frac{\widetilde{\alpha} _m}{m} \, e^{- i \, m \, \sigma} \right) e^{ i \, m \, \tau} \, .
\end{equation}
which is typical normal modes expansion for bosonic strings.
	
	Summarizing, the solution of Eq.~\eqref{EOM-f-1} is given by
	\begin{equation} \label{eq-sol}
\begin{split}
 \bm{X} (z, \sigma) = \bm{X}_0 - \frac{\sqrt{2 \, \alpha'} \, z^{2 \nu}}{2 \nu} \, \bm{\alpha} _0
  - i \, \sqrt{\frac{\alpha'}{2}} \sum _{m \neq 0}  \Big( \frac{\bm{\alpha} _m}{m} \, e^{i \, m\, \sigma} +  \frac{\widetilde{\bm{\alpha}} _m}{m} \, e^{- i \, m\, \sigma} \Big) \, \mathcal{E} (\nu, m \, ; \, z) \, .
\end{split}
\end{equation}
with $\bm{\alpha} _m ^\ast = \bm{\alpha} _{-m}$ and $\widetilde{\bm{\alpha}} _m ^\ast = \widetilde{\bm{\alpha}} _{-m}$, $\forall m \in \mathbb{Z}$.

	Now, it is easy to see that the constraints in Eq.~\eqref{Constraint-f-1} can be easily recast into the two following equations
	$$  
	|| \dot{\bm{X}} ||^2 + || \bm{X}' ||^2 = 0 \, , \qquad  \dot{\bm{X}} \cdot \bm{X}' = 0 \, .
	$$
Computing the derivatives of \eqref{eq-sol} one gets
\begin{equation} \label{Xdervdot}
\begin{split}
\dot{\bm{X}} = -\frac{\partial \bm{X}}{\partial z} &=  \sqrt{2 \, \alpha'} \, \, z^{2 \nu - 1} \, \bm{\alpha} _0\\
&- i \, z \, \sqrt{\frac{\alpha'}{2}} \sum _{m \neq 0} \texttt{Sgn} (m) \Big( \bm{\alpha} _m \, e^{i \, m\, \sigma} +  \widetilde{\bm{\alpha}} _m \, e^{- i \, m\, \sigma} \Big) \, \mathcal{E} (\nu - 1, m \, ; \, z) \, ,
\end{split}
\end{equation}
\be \label{Xdervprime}
\bm{X}' = 
- \, \sqrt{\frac{\alpha'}{2}} \sum _{m \neq 0}  \Big( \bm{\alpha} _m \, e^{i \, m\, \sigma} - \widetilde{\bm{\alpha}} _m \, e^{- i \, m\, \sigma} \Big) \, \mathcal{E} (\nu, m \, ; \, z) \, ,
\ee
with $\texttt{Sgn} (m)$ is the sign function and where we have taken profit of
\be 
\frac{\partial}{\partial z} \mathcal{E} (\nu, m \, ; \, z) = - |m| \, z \, \mathcal{E} (\nu - 1, m \, ; \, z) \, .
\ee

\begin{proof}
It is easy to see that \cite{Abram-Steg}
$$
\frac{\partial}{\partial z} \left[ z^\nu \, H_{- \nu} ^{(1)} (- m \, z) \right] = m \, z^\nu \, H_{- (\nu - 1)} ^{(1)} (- m \, z) \ ,
$$
$$
\frac{\partial}{\partial z} \left[ z^\nu \, H_{- \nu} ^{(2)} (m \, z) \right] = - m \, z^\nu \, H_{- (\nu - 1)} ^{(2)} (m \, z) \ ,
$$
therefore, from the definition in \eqref{new-function} one has that
\begin{equation*}
\begin{split}
\frac{\partial}{\partial z} \mathcal{E} (\nu, m \, ; \, z) &=
\sqrt{\frac{\pi}{2}} \, \frac{\partial}{\partial z} \Bigg[ z^\nu \times
\left\{
\begin{aligned}
& \sqrt{-m} \, H ^{(1)}_{-\nu} (- m z) \, , \,\, m \in \mathbb{Z} ^- \, , \\
& \sqrt{m} \, H ^{(2)}_{-\nu} (m z) \, , \,\, m \in \mathbb{N} \, ,  \\
 \end{aligned}
\right. \Bigg] \\
&= 
\sqrt{\frac{\pi}{2}} \,  z^\nu \times
\left\{
\begin{aligned}
& m \, \sqrt{-m} \, H ^{(1)}_{-(\nu - 1)} (- m z) \, , \,\, m \in \mathbb{Z} ^- \, , \\
& - m \, \sqrt{m} \, H ^{(2)}_{-(\nu - 1)} (m z) \, , \,\, m \in \mathbb{N} \, ,  \\
 \end{aligned}
\right. \\
&= 
- |m| \, z \, \Bigg[ \sqrt{\frac{\pi}{2}} \,  z^{\nu - 1} \times
\left\{
\begin{aligned}
& \sqrt{-m} \, H ^{(1)}_{-(\nu - 1)} (- m z) \, , \,\, m \in \mathbb{Z} ^- \, , \\
& \sqrt{m} \, H ^{(2)}_{-(\nu - 1)} (m z) \, , \,\, m \in \mathbb{N} \, ,  \\
 \end{aligned}
\right. \Bigg] =\\
&=
- |m| \, z \, \mathcal{E} (\nu - 1, m \, ; \, z) \, .
\end{split}
\end{equation*}
\end{proof}

	Then, if we now consider the expressions for $\dot{\bm{X}}$ and $\bm{X}'$, computed above, it is easy to see that the constraint equations do not lead, for the general $0 < \alpha < 1$, to the typical (ordinary) Virasoro constraints. Therefore, the study of the mass spectrum for the fractional bosonic string, at the quantum level, together with the formulation of a sort of generalized form of the of the classical Virasoro conditions appear to be more involved and they are, thus, left for future studies.  
\section{Hamiltonian for $f(\tau, \sigma) = 1$} \label{sec-H}
If we perform the Legendre transform of the \textit{fractional Lagrangian density}
\be 
\mathcal{L} (\tau , \sigma) = \frac{(t-\tau)^{\alpha-1}}{4\pi\alpha'\,\Gamma(\alpha)} \, \left[|| \bm{\dot{X}} ||^2- || \bm{X}' ||^2\right] \, ,
\ee
we get the Hamiltonian for the fractional bosonic string in the \textit{fractional conformal gauge}, \ie
\begin{eqnarray}
H(\tau) &=&\int_0^{2\pi}d\sigma\, \left\{ \dot{\bm{X}}\cdot \bm{\mathcal{P}}^\tau (\tau, \sigma) - \mathcal{L} (\tau , \sigma)\right\}\nonumber\\
&=&\frac{(t-\tau)^{\alpha-1}}{4\pi\alpha'\Gamma(\alpha)}\int_{0}^{2\pi}d\sigma \left(||\dot{\bm{X}}||^2+||\bm{X}'||^{2}\right),
\end{eqnarray}
where 
\be \label{canonical-momentum}
\bm{\mathcal{P}}^\tau (\tau , \sigma) = \frac{\partial \mathcal{L}  (\tau , \sigma)}{\partial \dot{\bm{X}}}  = 
\frac{(t - \tau)^{\alpha - 1}}{2 \, \pi \, \alpha' \, \Gamma (\alpha)} \, \dot{\bm{X}} (\tau , \sigma)
 \, .
\ee 
Using the constraints it is easy to see that $H=0$. It follows that the proper Hamiltonian is a given by
\begin{eqnarray}
H (\tau) = \int_0^{2\pi}d\sigma\, \left[
\lambda_1(\tau ,\sigma)\, \bm{\mathcal{P}}^\tau\cdot \bm{X}'+\lambda_2 (\tau,\sigma)\left( || \bm{\mathcal{P}}^\tau ||^2+\frac{(t - \tau)^{2\alpha-2}}{4\pi^2 \alpha'^2 \Gamma^2(\alpha)}\, || \bm{X}' ||^2\right)\right],
\end{eqnarray}
with $\lambda_1$ and $\lambda_2$ arbitrary functions of $z$ and $\sigma$.

	Using the Poison brackets 
\begin{eqnarray}
\left\lbrace X^{\mu}(\tau,\sigma),X^\nu(\tau,\sigma')\right\rbrace_{P.B.}&=&\left\lbrace\mathcal{P}^{\tau\mu}(\tau,\sigma),\mathcal{P}^{\tau\nu}(\tau,\sigma')\right\rbrace_{P.B.}=0\,;\\
\left\lbrace X^\mu(\tau,\sigma),\mathcal{P}^{\tau\nu}(\tau,\sigma')\right\rbrace_{P.B.}&=&\eta^{\mu\nu}\,\delta(\sigma-\sigma'),
\end{eqnarray}
and the fact that 
$$\left\lbrace H(\tau,\sigma),\mathcal{P}^{\tau\mu}(\tau,\sigma')\right\rbrace_{P.B.}=- \dot{\mathcal{P}}^{\tau\mu}(\tau,\sigma) \, ,$$
together with
$$\left\lbrace H (\tau,\sigma), X^{\mu}(\tau,\sigma')\right\rbrace_{P.B.}=-\dot{X} ^\mu (\tau,\sigma) \, ,$$
 one can infer the relations
\begin{eqnarray}
\dot{\bm X}(\tau,\sigma)&=&\lambda_1(\tau,\sigma)\,\bm{X}' (\tau,\sigma)+2\,\lambda_2(\tau,\sigma)\,\bm{\mathcal{P}}^{\tau}(\tau,\sigma) \, , \\
\dot{\bm{\mathcal{P}}}^{\tau}(\tau,\sigma)&=&\partial_\sigma \left[ \lambda_1(\tau,\sigma)\,\bm{\mathcal{P}}^\tau+2 \lambda_2(\tau,\sigma)\,\bm{X}'(\tau,\sigma)\right] \, .
\end{eqnarray}

	Now, If one sets 
	$$ 
	\lambda_1 (\tau,\sigma) =0 \, , \qquad 
	\lambda_2 (\tau,\sigma) = \frac{\pi \, \alpha' \, \Gamma (\alpha)}{(t - \tau)^{\alpha-1}} \, ,
	$$ 
	we recover the canonical momentum \eqref{canonical-momentum} and the equation of motion \eqref{EOM-f-1}. Therefore, this specific choice of $\lambda_1$ and $\lambda_2$ defines our \textbf{fractional conformal gauge}.  
	
	If we now take profit of the mode expansion \eqref{eq-sol} making $\tau=t-z$ and $\alpha=2-2\nu$, the Hamiltonian takes the form
\begin{eqnarray}\label{Hamilttau}
H (z) = \frac{1}{\Gamma(2-2\nu)}\,\left[L_0(\nu;z)+\tilde{L}_0(\nu;z)\right],
\end{eqnarray}
with the generalized Virasoro coefficients 
\begin{eqnarray}
L_0(\nu;z)\equiv\frac{1}{4} \sum _{m \in\mathbb{Z}}\left[ \bm{\alpha} _m\cdot \bm{\alpha} _{-m}\,\mathcal{G}_1(\nu,m;z)+2\,\bm{\alpha} _m\cdot \tilde{\bm{\alpha}} _{-m}\,\mathcal{G}_2(\nu,m;z)\right],\\
\tilde{L}_0(\nu;z)\equiv\frac{1}{4} \sum _{m\in\mathbb{Z}}\left[ \tilde{\bm{\alpha}}_m\cdot \tilde{\bm{\alpha}}_{-m}\,\mathcal{G}_1(\nu,m;z)+2\,\tilde{\bm{\alpha}} _m\cdot \bm{\alpha} _{-m}\,\mathcal{G}_2(\nu,m;z)\right].
\end{eqnarray} 
Here, we used the fact that $\bm{\alpha}_0=\tilde{\bm{\alpha}}_0$, and we have also defined $\mathcal{G}_1(\nu,0;z)=2$, $\mathcal{G}_2(\nu,0;z)=0$, and for $m\neq 0$
\begin{eqnarray}
\mathcal{G}_1(\nu,m;z)\equiv z^{1-2\nu}\left[ z^2\,\mathcal{E}(\nu - 1, m \, ; \, z) \,\mathcal{E} (\nu - 1, -m \, ; \, z)+\mathcal{E}(\nu, m \, ; \, z) \,\mathcal{E} (\nu, -m \, ; \, z) \right]\,,\\[2mm]
\mathcal{G}_2(\nu,m;z)\equiv z^{1-2\nu}\left[ z^2\,\mathcal{E}(\nu - 1, m \, ; \, z) \,\mathcal{E} (\nu - 1, -m \, ; \, z)-\mathcal{E}(\nu, m \, ; \, z) \,\mathcal{E} (\nu, -m \, ; \, z)\right]\, .
\end{eqnarray}
Now, it is easy to see that in the limit $\nu\rightarrow 1/2$, we get
\begin{eqnarray}
\mathcal{G}_1(1/2,m;z)=2\quad,\quad \mathcal{G}_2(1/2,m;z)=0\, ,
\end{eqnarray}
that leads to the well known result
\begin{eqnarray}
H&=&L_0+\tilde{L}_0\,,\nonumber\\
&=&\frac{1}{2}\sum _{m \in\mathbb{Z}}\left[ \bm{\alpha} _m\cdot \bm{\alpha} _{-m}+\tilde{\bm{\alpha}}_m\cdot\tilde{\bm{\alpha}} _{-m}\right].
\end{eqnarray}
%
%
%%%%%
\section{Fractional light-cone gauge} \label{sec-LC}

	In the usual formalism of the bosonic strings, the Virasoro constraints arise from residual symmetries of the system after fixing the conformal gauge (see \eg \cite{lust, Zwiebach}). Here, we also have some variations of the same constraints. Therefore, it is natural to think that we might used those residual symmetries in order to set
	\be \label{resdgauge}
	n_\mu\,X^\mu(\tau,\sigma)\propto g(\tau,\alpha)
	\ee
with $n_\mu$ a constant vector and where $g(\tau,\alpha)$ is such that in the limit $\alpha\rightarrow 1$ we have $g(\tau,1)\propto \tau$.  Now, selecting the light-cone gauge consist in imposing \eqref{resdgauge} with a constant vector $n_\mu$ that gives $n_\mu\,X^\mu(\tau,\sigma)=X^+(\tau,\sigma)$, \ie with $n_\mu=\frac{1}{\sqrt{2}}(1,0,\ldots,0,1)$. 

	In other terms, introducing the light-cone coordinates, \ie
$$ X^{\pm} \equiv \frac{X^{0} \pm X^{d-1}}{\sqrt{2}} \, , $$
that allows us to rewrite a vector in the Minkovsky space as $\bm{X} = ( X^{+} , \, X^{-} \, , X^{1} , \ldots, X^{d-2} )$, then a fractional version of light-cone gauge corresponds to $X^+ (\tau, \sigma) \propto g(\tau,\alpha)$.
	
	Therefore, denoting with
	$$ \bm{p} = \int_0^{2\pi}d\sigma\,\bm{\mathcal{P}}^{\tau} \, , $$
	then we can define \textit{the fractional light-cone gauge} as
	\be\label{Xplus}
	X^+(z)=-\frac{\sqrt{2\,\alpha'}}{2\nu} \,\alpha_0^+\,z^{2\nu},
	\ee
where $\alpha_0^+=\sqrt{\frac{\alpha'}{2}}\,\Gamma(2-2\nu)\,p^+$ with $p^+\neq0$. The main use of this gauge is usually related to the reduction of the constraints to a simpler mathematical form. Indeed, taking profit of the dependence on $z$ of $X^+$, the constrains take the form
	\begin{eqnarray}
	|| \dot{\bm{X}} ||^2 + || \bm{X}' ||^2 = 0 \,\, &\rightsquigarrow& \,\, 
	2\,\dot{X}^+\,\dot{X}^- =\dot{X}^i\,\dot{X}^i+X'^i\,X'^i\, \,,\\
	\dot{\bm{X}}\cdot\bm{X}'=0\,\, &\rightsquigarrow& \,\, 
	\dot{X}^+\,X'^- =\dot{X}^i\,X'^i\,,
			\end{eqnarray}
where the sum over repeated indices (in this case $i=1, \ldots,d-2$) is omitted, and we can write
	\begin{eqnarray}\label{Xdotconst}
	\dot{X}^- &=&\frac{\Gamma(2-2\nu)\,z^{1-2\nu}}{2\,\alpha'\,p^+}\left[\dot{X}^i\,\dot{X}^i+X'^i\,X'^i\right]\, ,\\[2mm]\label{Xprimeconst}
	X'^- &=&\frac{\Gamma(2-2\nu)\,z^{1-2\nu}}{\alpha'\,p^+}\,\dot{X}^i\,X'^i\,,
			\end{eqnarray}
Thus, one can see that $X^-$ is completely determined by $p^+$ and the tranverse directions $X^i$.

\subsection{Hamiltonian and masses}
	We have already computed the Hamiltonian in \eqref{Hamilttau}, which in the fractional light-cone coordinates can be written as
\begin{eqnarray}
H (\tau)&=&\alpha'\,p^+\,p^-\,(t-\tau)^{1-\alpha}\nonumber\\
&=&\frac{1}{\Gamma(2-2\nu)}\,\left[L_0(\nu;z)+\tilde{L}_0(\nu;z)\right],
\end{eqnarray}
with $\bm{\alpha}_m (\bm{\tilde{\alpha}}_m)$ that ultimately reduce to $\alpha^i_{m}(\tilde{\alpha}^i_{m})$, and 
\begin{eqnarray}
p^-&=&\int_0^{2\pi}d\sigma\,\mathcal{P}^{\tau -}\nonumber\\
&=&\frac{(t-\tau)^{2\alpha-2}}{4\,\pi\,\alpha'^2\,p^+}\,\int_0^{2\pi}d\sigma\, \left[\dot{X}^i\,\dot{X}^i+X'^i\,X'^i\right].
\end{eqnarray}

	Then, from the mass-shell condition we can define the mass of a fractional bosonic string, which is given by
\be 
M^2 = - || \bm{p} ||^2 \, .
\ee 
Using the fractional light-cone gauge and the above equation, we obtain 
\begin{eqnarray}
M^2(\tau)= \frac{2 \, (t - \tau)^{1-2\nu}}{\alpha'}\,H (\tau) - p^i\,p^i,
\end{eqnarray}
that can be rewritten as
\begin{eqnarray}
\frac{\alpha'}{2}\,M^2(\tau)&=&\frac{(t - \tau)^{1-2\nu}}{\Gamma(2-2\nu)}\,\left[L_0(\nu; \tau)+\tilde{L}_0(\nu; \tau)\right]-\frac{\alpha'}{2} \,p^i\,p^i.
\end{eqnarray}
Now, if we take the limit $\nu\rightarrow1/2$ we recover the classical result
\begin{eqnarray}
\frac{\alpha'}{2}\,M^2 = \frac{1}{2}\sum _{m\neq0}\left[ \bm{\alpha} _m\cdot \bm{\alpha} _{-m}+\tilde{\bm{\alpha}}_m\cdot\tilde{\bm{\alpha}} _{-m}\right].
\end{eqnarray}

\section{Conclusions} \label{sec-conclusions}

	After reviewing some aspect of fractional variational problems we have presented a quite general fractional modification of the Polyakov action and we have discussed the corresponding equations of motion, together with the connection with an extended notion of Nambu-Goto action.

	Then, we have simplified the problem by considering a fractionalization of the action with respect to the parameter $\tau$ alone. For this simplified action we have then computed the corresponding equations of motion and we have studied the underlying symmetries of the theory.
	
Unfortunately, due to the absence of the $\tau$-reparametrization invariance, we are still left with an extra degree of freedom that results in an extra functional dependence on the WS parametrization in the induced metric $h_{ab}$. This makes any attempt for a general solution of the equations of motion hopeless. In order to avoid this problem, we consider a very specific realization of the model by setting $f(\tau, \sigma) = 1$.

Then, we have computed the hamiltonian function for our model and we have also provide a characterization for the fractional conformal gauge.

Finally, we have discussed the notion of fractional light-cone gauge and we have derived the classical mass for a fractional bosonic string.
	 
	A precise study of the residual symmetries as well as the of the quantization of the fractional bosonic string, even in the simplified setting, appear to be some rather involved problems and are therefore left for future studies.

	To conclude on a more physical note, on basic phenomenological principles the natural extension of this work is the inclusion of fermionic degrees of freedom to the theory, in order to fill the baryonic components in the Universe. The insertion of these new degrees of freedom might be performed by adding WS supersymmetry like in the usual Ramond-Neveu-Schwarz spinning string (RNS formalism). Furthermore, it would be interesting to study how the superconformal gauge changes in order to accommodate these new objects.
	
%%%%%%%%%%%%%%%%
%%%% Acknowledgments
\section*{Acknowledgments}
This research was partially supported by INFN, research initiatives 
FLAG (A.G.) and ST$\&$FI (V.A.D.). Moreover, the work of A.G. has been carried out in the framework 
of GNFM and INdAM and the COST action Cantata.


\begin{thebibliography}{99}

\bibitem{Baleanu}
D.~Baleanu, S.~Muslih,
\emph{Physica~Scripta} \ {\bf 72}, 119 (2005).

\bibitem{MainardiB}
F.~Mainardi, {\em Fractional Calculus and Waves in Linear Viscoelasticity}, 
Imperial College Press \& World Scientific, London -- Singapore, 2010.

\bibitem{IC-AG-FM-ZAMP}
I.~Colombaro, A.~Giusti, F.~Mainardi, 
{\em Z.~Angew.~Math.~Phys.} \ \textbf{68}, 62 (2017). 
   
\bibitem{IC-AG-FM-Bessel}
I.~Colombaro, A.~Giusti, F.~Mainardi, 
{\em Meccanica} \ \textbf{52}, 825 (2017).

\bibitem{Fabrizio}
M.~Fabrizio,
{\em Fract.~Calc.~Appl.~Anal.} \ {\bf 18}, 1074 (2015).

\bibitem{AG-FCAA-2017}
A.~Giusti,
{\em Fract.~Calc.~Appl.~Anal.} \ {\bf 20}, 854 (2017).

\bibitem{extra}
A.~Giusti, F.~Mainardi, 
{\em Eur.~Phys.~J.~Plus} \ \textbf{131}, 206 (2016). 

\bibitem{AG-FM_MECC16}
A.~Giusti, F.~Mainardi, 
{\em Mecanica} \ \textbf{51}, 2321 (2016).

\bibitem{JMP-mio}
A.~Giusti, 
{\em J.~Math.~Phys.} \ {\bf 59}, 013506 (2018).

\bibitem{Garra}
P.~Artale~Harris, R.~Garra,
{\em J.~Math.~Phys.} \ {\bf 58}, 063501 (2017).  

\bibitem{Silvia-1}
S.~Vitali, G.~Castellani, F.~Mainardi,
\emph{Chaos~Solitons~$\&$~Fractals} \ \textbf{102}, 467 (2017).

\bibitem{Vacaru}
  S.~I.~Vacaru,
  \emph{Int.~J.~Theor.~Phys.} \ {\bf 51}, 1338 (2012).

\bibitem{Ata}
T.~M.~Atanackovic, S.~Konjik, S.~Pilipovic,
\emph{	J.~Phys.~A:~Math.~Theor.} \ {\bf 41}, 095201 (2008).

\bibitem{Almeida}
R.~Almeida, S.~Pooseh, D.~F.~M.~Torres,
\emph{Nonlinear~Analysis} \ {\bf 75}, 1009 (2012).

\bibitem{Torres-book}
A.B.~Malinowska, D.F.M.~Torres,
\emph{Introduction to the fractional calculus of variations},
World Scientific Publishing Company, 2012.

\bibitem{El}
R.A.~El-Nabulsi, D.F.M.~Torres, 
\emph{J.~Math.~Phys.} \ {\bf 49}, 053521 (2008).

\bibitem{Gorenflo-Mainardi}
R.~Gorenflo, F.~Mainardi,
Fractional Calculus: Integral and Differential Equations of Fractional Order,
in A. Carpinteri and F. Mainardi (Editors): \emph{Fractals and Fractional Calculus in Continuum Mechanics}, Springer Verlag, Wien and New York, 1997; pg. 223.

\bibitem{Kilbas}
A.~A.~Kilbas, H.~M.~Srivastava, J.~J.~Trujillo, 
{\em Theory and Applications of Fractional Differential Equations}, 
Elsevier, Boston, 2006.

\bibitem{Calcagni-rev}
G.~Calcagni,
\emph{Adv.~Theor.~Math.~Phys.} \ {\bf 16}, 549 (2012).

\bibitem{Calcagni-PRL}
G.~Calcagni, 
\emph{Phys.~Rev.~Lett.} \ {\bf 104}, 251301 (2010).

\bibitem{Cosmo-1}
 V.~K.~Shchigolev, 
 \emph{Mod.~Phys.~Lett.~A} \ \textbf{28}, 1350056 (2013).

\bibitem{Cosmo-2}
V.~K.~Shchigolev,
\emph{Eur.~Phys.~J.~Plus} \ \textbf{131}, 256 (2016)

\bibitem{WSEAS}
C.~Udriste, D.~Opris, 
\emph{WSEAS~Trans.~Math.} \ {\bf 7}, 19 (2008).

\bibitem{deser}
  S.~Deser, B.~Zumino,
  \textit{Phys.~Lett.} \  {\bf 65B}, 369 (1976).

\bibitem{brink}
  L.~Brink, P.~Di Vecchia, P.~S.~Howe,
  \textit{Phys.\ Lett.} \ {\bf 65B}, 471 (1976).
 
\bibitem{polyakov}
   A.~M.~Polyakov,
  \textit{Phys.\ Lett.} \  {\bf 103B}, 207 (1981).
  
\bibitem{lust}
R.~Blumenhagen, D.~L\"{u}st and S.~Theisen,
\emph{Basic Concepts of String Theory}, 
Springer Science $\&$ Business Media, 2012.

\bibitem{Zwiebach}
B.~Zwiebach,
\emph{A first course in string theory}, 
Cambridge university press, 2004.

\bibitem{Fractals}
F.B.~Tatom,
\emph{Fractals} \ {\bf 3}, 217 (1995).

\bibitem{Abram-Steg}
M. Abramowitz and I. A. Stegun, 
\textit{Handbook of Mathematical Functions, with Formulas, Graphs, and Mathematical Tables},
Dover, 1972.

%\bibitem{a}
%Author, \emph{Title}, \emph{J. Abbrev.} {\bf vol} (year) pg.

%\bibitem{b}
%Author, \emph{Title},
%arxiv:1234.5678.

%\bibitem{c}
%Author, \emph{Title},
%Publisher (year).

\end{thebibliography}
\end{document}